\documentclass[preprint2]{aastex}

\usepackage{amssymb}  % extended symbols lib
\usepackage{multirow}  % muti-row table cells
\usepackage{amsmath}  % extended equations lib (split)
\usepackage{mathrsfs} % extended math fonts (mathscr)
\usepackage{paralist} % inline enumeration (for Table ref lists)
\usepackage[usenames]{color}  % colored text

\newcommand{\TNSN}{TN\,SN}
\newcommand{\TNSNe}{TN\,SNe}

\newcommand{\Ni}{\ensuremath{^{56}\mbox{Ni}}}

\newcommand{\Av}{{\ensuremath{\mbox{A$_{V}$}}}}

\newcommand{\tpk}{{\ensuremath{\mbox{t{$_{\mbox{\tiny pk}}$}}}}}
\newcommand{\mue}{\ensuremath{\mu_{e}}}

\newcommand{\bftheta}{\ensuremath{\mbox{\boldmath$\theta$}}}
\newcommand{\bfvartheta}{\ensuremath{\mbox{\boldmath$\vartheta$}}}
\newcommand{\bfPhi}{\ensuremath{{\bf \Phi}}}
\newcommand{\bfD}{\ensuremath{{\bf D}}}

\newcommand{\Om}{\ensuremath{\Omega_{\mbox{\scriptsize M}}}}
\newcommand{\Ok}{\ensuremath{\Omega_{\mbox{\scriptsize k}}}}
\newcommand{\Ode}{\ensuremath{\Omega_{\mbox{\scriptsize DE}}}}

\shorttitle{Fuzzy Supernova Estimation}
\shortauthors{Rodney and Tonry}
\begin{document}

\title{Fuzzy Supernova Templates II: Parameter Estimation}

\author{Steven A. Rodney}
\affil{Department of Physics and Astronomy,Johns Hopkins University\\
 Baltimore, MD 21218}
\email{srodney@jhu.edu}
%\and
\author{John L. Tonry}
\affil{Institute for Astronomy, University of Hawaii\\
Honolulu, HI 96822}
\email{jt@ifa.hawaii.edu}

\begin{abstract}
Wide field surveys will soon be discovering Type Ia supernovae (SNe) at
rates of several thousand per year.  Spectroscopic follow-up can only
scratch the surface for such enormous samples, so these extensive data
sets will only be useful to the extent that they can be characterized
by the survey photometry alone.  In a companion paper (Rodney and
Tonry, 2009) we introduced the SOFT method for analyzing SNe using
direct comparison to template light curves, and demonstrated its
application for photometric SN classification. In this work we extend
the SOFT method to derive estimates of redshift and luminosity
distance for Type Ia SNe, using light curves from the SDSS and SNLS
surveys as a validation set. 
Redshifts determined by SOFT using light
curves alone are consistent with spectroscopic redshifts, showing a
root-mean-square scatter in the residuals of $RMS_z=0.051$. 
%Comparing the SOFT redshift error distribution against galaxy photo-z
%errors from $ugriz$ photometry, we find that the two methods are
%complementary: SOFT can weed out bad host galaxy photo-z's,
%strengthen the good ones, and provide a reliable photometric redshift
%estimate even for host-less SNe.
SOFT can also derive simultaneous redshift and distance estimates,
yielding results that are consistent with the currently favored
$\Lambda$CDM cosmological model.  When SOFT is given spectroscopic
information for SN classification and redshift priors, the RMS scatter
in Hubble diagram residuals is 0.18 mags for the SDSS data and 0.28
mags for the SNLS objects.  Without access to any spectroscopic
information, and even without any redshift priors from host galaxy
photometry, SOFT can still measure reliable redshifts and distances,
with an increase in the Hubble residuals to 0.37 mags for the combined
SDSS and SNLS data set.  Using Monte Carlo simulations we predict that
SOFT will be able to improve constraints on time-variable dark energy
models by a factor of 2--3 with each new generation of large-scale SN
surveys.
\end{abstract}

The use of Thermonuclear Supernovae (\TNSNe, i.e. Type Ia SNe) as
cosmological standard candles requires the determination of each SN's
redshift and distance modulus.  This has traditionally been done with
two parallel measurements: a spectrum confirms the object's class and
provides a very accurate redshift, while a broad band light curve
allows the estimation of luminosity distance.  Technological advances
in wide-field imagers are now opening the door for a new generation of
large area surveys. Projects such as the Palomar Transient Factory
\citep[PTF;][]{Rau:2009,Law:2009}, and Pan-STARRS 1
\citep[PS1;][]{Kaiser:2002} are already beginning to amass SN samples
that will eventually number in the thousands.  In coming years
Pan-STARRS 4\footnote{\url{http://pan-starrs.ifa.hawaii.edu}},
LSST\footnote{\url{http://www.lsst.org}}, and
JDEM\footnote{\url{http://nasascience.nasa.gov/missions/jdem}} will
uncover tens of thousands of SNe.  From these vast collections of SN
light curves, only a small fraction will be spectroscopically
observed, meaning that any analyses applied to them will have to rely
on photometry alone.

In \citet{Rodney:2009} (hereafter Paper~I), we introduced SOFT
(Supernova Ontology with Fuzzy Templates) and described in detail the
structure and form of the SOFT light curve analysis method.  In this
paper, Section \ref{sec:LightCurveModels} begins with a brief review
of the basis for the SOFT method and the fuzzy set theory framework
for light curve comparisons.  In Section \ref{sec:ParameterEstimation}
we apply the SOFT models to the task of estimating cosmological
parameters using SN light curves. Verification tests using data from
the Sloan Digital Sky Survey (SDSS) and the Supernova Legacy Survey
(SNLS) are described in Section \ref{sec:VerificationTests}. The
results of these tests are then extended to produce Monte Carlo
simulations of future surveys to examine how well an approach such as
SOFT will be able to constrain dark energy models.  A discussion of
the inherent biases in our template estimators is provided in Section
\ref{sec:BiasCorrections}, and a summary of results is presented in
Section \ref{sec:Summary}.

\section{Light Curve Models}
\label{sec:LightCurveModels}

The shape of an observed SN light curve can be described by two sets
of parameters. The first set, which we denote \bfPhi, consists of all
relevant physical parameters that determine the intrinsic shape and
color of the light curve (e.g. the \Ni\ mass, the degree of mixing in
the stellar interior, etc.).  A second set of parameters, \bftheta,
relate to the location of the SN in space and time: redshift $z$,
luminosity distance \mue,\footnote{The SOFT luminosity distance
  parameter \mue\ is the distance modulus residual relative to an
  empty universe.  See equation 3 of Paper I.}  host galaxy extinction
\Av, and time of peak \tpk.  These four {\em location parameters}
modify the intrinsic light curve shape by stretching and dimming it
into the form that we eventually observe.

We do not have a complete physical model for SN explosions that could
efficiently and reliably translate a vector of physical parameters
\bfPhi\ into a precise light curve prediction.  Therefore, most
\TNSN\ light curve fitters -- such as MLCS \citep{Riess:1996,Jha:2002}
and SALT \citep{Guy:2005,Guy:2007} -- use one or more paramaters
(e.g. $\Delta$, or X1 and C) to define the shape and color of a
broad-band SN light curve.
By contrast, the SOFT method has no shape
or color parameters, but instead uses a collection of SN light curve
templates to provide examples of possible shapes and colors.
These template light curves are then warped to reproduce the effects
of a change in location \bftheta.  By comparing this adapted light curve
model against a candidate SN object, we can derive the likelihood that
the candidate has an intrinsic light curve shape similar to the
template, and is at the assumed location \bftheta. 

In Paper I we discussed how our template-based approach is best served
by utilizing the framework of fuzzy set theory \citep{Zadeh:1965}, and
we applied SOFT to the task of SN classification.  In this paper we
will now explore how the application of fuzzy logic allows us to
extend the SOFT method to the problem of estimating cosmologically
useful parameters such as a SN's redshift and distance.

\section{Parameter Estimation}
\label{sec:ParameterEstimation}

Suppose we have a candidate object called SN~X whose observed light
curve is given by a series of points $D_i=(t_i, f_i, \sigma_i)$.  Here
$t_i$ is the observation time, $f_i$ is the measured flux, and
$\sigma_i$ is the uncertainty for epoch $i$.  The
complete light curve with N data points is denoted by the vector \bfD.
If we assume a set of location parameters \bftheta\ then a light
curve model $M_j$ that realizes a given \bfPhi\ provides a prediction
for a SN's flux as a function of time: $\mathcal{F}_j(t_i,\bftheta)$.

In the framework of traditional probability theory, we
 would describe the probability of observing the
data \bfD\ given model $M_j$ and location \bftheta\ as the combined
probability of the data-minus-model error terms: 

\begin{equation}
  p(\bfD|\bftheta,M_j) =  \prod\limits_{i=1}^N
  \frac{1}{\sqrt{2\pi}\sigma_i}
  \mbox{exp}\left(\frac{-(f_i-\mathcal{F}_{j}(t_i,\bftheta))^2}{2\sigma_i^2}\right) 
\label{eqn:p(D|theta,Mj)}
\end{equation}

\noindent (see Equations 4-6 in Paper I).  We then compute the
probability that our candidate object resides at any particular
location \bftheta\ by applying Bayes theorem:

\begin{equation}
  p(\bftheta|\bfD,M_j) =   \frac{ p(\bftheta|M_j)~ 
    p(\bfD|\bftheta,M_j)}{ \int\limits_\theta{ p(\bftheta|M_j)~ 
    p(\bfD|\bftheta,M_j)~~d\bftheta} }
\label{eqn:p(theta|D,Mj)}
\end{equation}
  
\noindent This gives us a probability distribution over the four
dimensions of \bftheta, under the assumption that our candidate
SN~X is physically identical to the model $M_j$.   
Variations of this approach using parameterized light curve models
have been applied to SN classification
\citep{Sullivan:2006b,Johnson:2006,Kuznetsova:2007,Poznanski:2007a} as
well as to the estimation of SN redshifts
\citep{Barris:2004,Kim:2007,Gong:2010}. 

With SOFT, however, we remove the need for a parametric model of the
SN light curve by limiting ourselves to a finite set of fixed-shape
models. To keep within the framework of traditional probability
theory, we could apply Equation~\ref{eqn:p(theta|D,Mj)} to derive
probabilities from all our models and then combine the results using a
method such as Bayesian Model Averaging \citep{Hoeting:1999}.
Unfortunately, that approach is applicable only in the case where our
set of models is complete and non-redundant.  If either of those two
conditions is not satisfied, then the normalization inherent in
applying Bayes' theorem is not valid.  In Paper I we have argued that
it is impossible to completely satisfy both conditions when using a
finite library of fixed-shape templates.

\subsection{Biased Estimators}
\label{sec:BiasedEstimators}

In addition to problems of logical incongruence described above and in
Paper I, our finite set of templates will necessarily introduce an
element of bias to any parameter estimates.  To illustrate this
problem, suppose we have an extremely sparse template library,
consisting of only two template models, $M_1$ and $M_2$.
%, which have
%underlying physical parameters \bfPhi$_1$ and \bfPhi$_2$.  
Now let us observe a candidate object, SN~X, which has physical
parameters \bfPhi$_X$ that are intermediate between \bfPhi$_1$ and
\bfPhi$_2$ (e.g., perhaps the candidate has a \Ni\ mass that is
bracketed by the two templates: $M_{Ni,1} < M_{Ni,x} < M_{Ni,2}$).
SN~X is not a perfect match with either of the two models, so applying
Eq.~\ref{eqn:p(theta|D,Mj)} will give us a biased parameter estimate
from each of them.  
For simplicity suppose that all the location
parameters are fixed except for the cosmological dimming,
\bftheta$=$\mue.  The model $M_2$ with a higher $M_{Ni}$ will have a
higher bolometric flux than SN~X \citep{Arnett:1982}.  Thus, the best
match from the $M_2$ model will have to increase the cosmological
dimming \mue\ away from the true value to compensate.\footnote{Recall
  that \mue\ is a distance measured in magnitudes, so a higher
  \mue\ makes the intrinsically bright model $M_2$ appear farther away
  and fainter, bringing it closer in form to the less luminous SN~X.}
Conversely, the fainter $M_1$ model must decrease the cosmological
dimming (lower \mue\,$\rightarrow$\,higher flux) to bring it up to the
level of SN~X.  Therefore $M_1$ will report a lower \mue\ as the most
likely parameter.
%If we were to use the Bayesian Model Averaging
%approach in this circumstance, we would sum together the two
%probability distributions, weighted by their PMP's, using
%Eq.~\ref{eqn:p(theta|D)}, and we would get a composite probability
%distribution that is bimodal, with a peak from $M_2$ biased toward
%high \mue\ values and a peak from $M_1$ biased toward low
%\mue\ values.

\begin{figure*}[tb]
  \centering
  \includegraphics[draft=False,width=\textwidth]{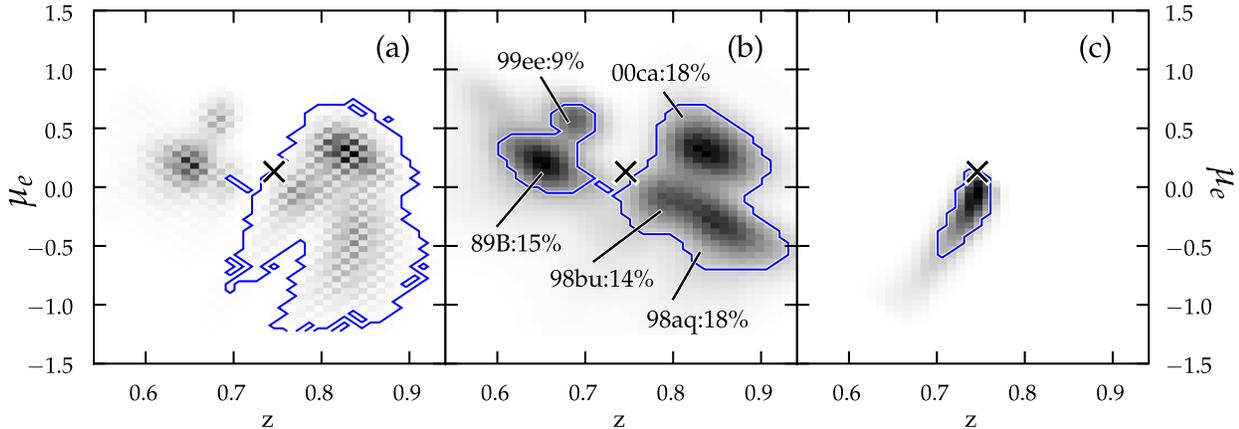}
  \caption{ Demonstration of the parameter estimation biases inherent
    in a finite set of light curve templates, and the bias averaging
    enabled by fuzzy templates.  The candidate SN used here was the
    SNLS object 04D2ja.  {\bf(a)}~Probability distributions
    constructed using the strict Bayesian approach of
    Equations~\ref{eqn:p(D|theta,Mj)} and \ref{eqn:p(theta|D,Mj)}
    (i.e. without any model fuzziness).  The `x' marks the
    spectroscopic redshift of this SN and the corresponding distance
    modulus for a universe with
    $H_0,\Omega_M,\Omega_\Lambda=(70,0.3,0.7)$.  The gray-scale
    shading defines the probability distribution, and the contour line
    encloses the region about the peak that contains 68\% of the
    probability.  {\bf(b)}~The introduction of model fuzziness using
    $\sigma'_j$ as in Section \ref{sec:FuzzySNModels} expands the area
    of parameter space in which each model can provide a fair
    match. For this candidate SN, our template library finds 5
    templates that provide reasonable fits.  Here we display the
    ``probability'' cloud that defines each individual fuzzy set
    membership function, simply summing them together without using a
    proper fuzzy set combination operator. Each cloud is marked with
    the designation of its generating template and a percentage that
    indicates its integrated membership grade $\gamma$
    (Eq.~\ref{eqn:gammaj}).  {\bf(c)}~After rejecting outliers with a
    membership cut at 5\%, we combine the remaining fuzzy membership
    functions using the Dombi intersection operator.  This ``Fuzzy
    AND'' operation tends to average out the biases of any individual
    template.  }
  \label{fig:fuzzyANDdemo}
\end{figure*}

In Figure~\ref{fig:fuzzyANDdemo}a we illustrate this parameter bias
using a real SN light curve from the SNLS, the Type Ia SN~04D2ja.  We
compared this object's light curve against all 28 \TNSN\ templates, in
each case using equation \ref{eqn:p(theta|D,Mj)} to construct a
probability distribution in the z-\mue\ plane.  The sum of these
distributions, shown in Figure~\ref{fig:fuzzyANDdemo}a, is diffuse and
inaccurate, with multiple peaks spread across the parameter space.  No
individual template provides a clear match to the shape of the
candidate light curve, so each template requires an offset from the
true location in order to compensate for the shape mismatch.  
%The strict Bayesian interpretation of this probability distribution is
%that the candidate has a low likelihood of being located just about
%anywhere in the z-\mue\ plane.

\subsection{Fuzzy SN Models}
\label{sec:FuzzySNModels}

We now have two problems arising from the traditional (Bayesian)
approach to parameter estimation when using a finite set of templates.
First is the problem of violating the assumptions of completeness and
non-redundancy (see section 4 of Paper I), and second is the issue of
parameter estimation biases described above.  We propose that both of
these problems are resolved by invoking {\em fuzzy logic}, an
alternative logical framework that is better suited to the case of our
limited template library.

In Paper I we developed the basic structure of a fuzzy SN model.  Each
template model $M_j$ carries an associated ``model fuzziness'' term,
$\sigma'_j$, which is used to define a fuzzy set around that template.
The $\sigma'_j$ term quantifies how membership in that model's fuzzy
set falls off as light curves become more and more dissimilar to the
template curve:   
%The membership function is defined in a manner
%that is consistent with the structure of probability theory.  

%We then define a fuzzy set for the template model $M_j$ at location
%\bftheta\   and are located at position \bftheta.''\footnote{Note that the
%fuzziness of this set comes only from the ``similar to model $M_j$''
%phrase. The requirement to be ``located at position \bftheta'' by
%itself defines a crisp set.}
%As before, the model $M_j$ provides a prediction of the flux as a
%function of time for each possible location:
%$\mathcal{F}_j(t,\bftheta)$.  However, we can now explain the
%difference in flux between observations and model using a combination
%of the observational uncertainty $\sigma_i$ and the model fuzziness
%$\sigma'_j$:

\begin{equation}
  \begin{array}{ll}
  p''(\bfD|\bftheta,M_j) = & 
  \prod\limits_{i=1}^N{\frac{1}{\sqrt{2\pi}\sqrt{ \sigma_i^2 + {\sigma'_j}^2}}}\\
    & \times\ \mbox{exp}\left(\frac{-\left(f_i - \mathcal{F}_{j}(t_i,\theta)\right)^2}{ 2~(\sigma_i^2 + {\sigma'_j}^2)}\right)\\
  \end{array}
\label{eqn:p''(D|theta,Mj)}
\end{equation}

\noindent(cf. Eq.~\ref{eqn:p(D|theta,Mj)} and Paper I, Eq.~13).  

It is no accident that this expression for the likelihood is
functionally identical to what we would use in traditional probability
theory if $\sigma'_j$ was a simple model uncertainty.  There is a
conceptual difference between the two types of uncertainty: our model
fuzziness represents an intentional ``vagueness'' or
``indefiniteness,'' whereas the traditional model uncertainty
encapsulates an ``imprecision'' or ``inaccuracy.''  In the language of
set theory, our fuzzy model defines a set that includes ``All SNe that
are physically {\em similar} to model $M_j$.''  This leaves room for
our model to be applied to objects that we know to be slightly
different from our template.  By contrast, in traditional probability
theory a model $M_j$ defines a set that includes ``All SNe that are
physically {\em identical} to model $M_j$,'' and the addition of model
uncertainty would simply acknowledge that we don't know exactly what
our model $M_j$ really looks like.  Although the mathematical form is
the same, the fuzzy concept of vagueness allows us to adjust the
strength of $\sigma'_j$ to suit our purposes (increasing $\sigma'_j$
makes the model more inclusive but less predictive), whereas a
traditional model uncertainty should be derived strictly from
empirical evidence.  As we will see in the next section, this model
fuzziness also leads us to a powerful advantage within fuzzy logic,
namely that we can apply a fuzzy AND combination instead of the less
informative OR.

To convert the likelihood of Equation~\ref{eqn:p''(D|theta,Mj)} into a fuzzy membership grade 
we introduce a prior on the location parameters, $p(\bftheta|M_j)$, as 
well as a prior for the particular template in question, $p(M_j)$.
The final membership grade for our candidate SN~X in the
(\bftheta,$M_j$) fuzzy set is given by:  

\begin{equation}
  g(\bfD|\bftheta,M_j) = p(\bftheta|M_j)~ p(M_j)~ p^{\prime\prime}(\bfD|\bftheta,M_j)
  \label{eqn:g(D|theta,Mj)}
\end{equation}

\noindent(see Paper I, Eq.~14 and section 5.1 for a full
description of fuzzy membership functions).

To compare the quality of fit between two or more models, we define a
scalar quantity $\gamma$ as a
normalized integral over the location parameters \bftheta. For a
single model $M_j$ we have: 

\begin{equation}
  \gamma_j = \frac{\int\limits_\theta{g(\bfD|\bftheta,M_j)}\delta\theta}
  {\sum\limits_k~\int\limits_\theta{g(\bfD|\bftheta,M_k)}\delta\theta}
  \label{eqn:gammaj}
\end{equation}

\noindent where the summation is over all available models.
The value of $\gamma_j$ lies in the range [0,1] and measures
the candidate object's strength of association with model $M_j$
relative to all other models. This term can serve as the fuzzy analog
for a Bayesian posterior probability.
%When the $PMG_j$ value is very low, it
%indicates that the model under consideration has a substantially
%different light curve shape, implying a substantially different set of
%physical parameters.  

By adding an element of fuzziness to the SN models, we have alleviated
the logical inconsistencies of the Bayesian approach when applied to
our finite template library (see section 4.1 of Paper I). We have not,
however, removed the inherent biases in the parameter estimates of
individual templates.  Returning to our example from the
SNLS, in Figure~\ref{fig:fuzzyANDdemo}b we show a stack of membership
functions from comparison of the candidate SN against our fuzzy
\TNSN\ models. 
%Compared to Figure~\ref{fig:fuzzyANDdemo}a we see
%that the probability distributions from each template are smoothed
%out and still spread over a large region of parameter space. 
%Each
%template still provides a probability distribution (now called a
%membership function) that is biased by the mismatch of light curve
%shape between our candidate and the template. 
%However, these biased
%distributions have greater overlap, blurring the distinctions between
%peaks.  %In Figure~\ref{fig:fuzzyANDdemo}b 
The five most prominent membership function clouds are labeled with
the templates they originate from as well as a percentage showing the
value of $\gamma$ from Equation~\ref{eqn:gammaj}.
The addition of model fuzziness has smoothed the peaks of the
distribution, but individual membership functions are still biased by
the mismatch of light curve shape between our candidate and the
templates.
%Having established an independent membership function $g_j$ for each
%model $M_j$, we can derive parameter estimates for the true location
%of SN~04D2ja from each template.  Recognizing that the parameter
%estimates from individual $g_j$ distributions still carry inherent
%biases, 
We would like to be able to combine these membership functions
together in a way that gives us an improved {\em composite} parameter
estimate.  In the following section we explore how this can be
achieved using the structures of fuzzy logic.

\subsection{Fuzzy Combination}
\label{sec:FuzzyCombination}

The two principle operators for fuzzy logic can be defined in a number
of ways.  As described in the companion paper, we have adopted the
\citet{Dombi:1982} operators (see Paper I, Section 5.2, Equations 16
and 17), which define how two fuzzy membership functions can be
combined in a logical intersection (fuzzy AND) or union (fuzzy OR).
To choose whether to use the intersection or the union operator for
our SN models we must consider how much interaction between the
fuzzy sets is appropriate for the problem.

Suppose we have done two model comparisons, using models $M_1$ and
$M_2$ to get two membership functions $g_1=g(\bfD|\bftheta,M_1)$ and
$g_2=g(\bfD|\bftheta,M_2)$.  Let us also suppose that these two models
are based on SNe that have intrinsically very similar physical
characteristics: $\bfPhi_1 \sim \bfPhi_2$.  If it happens that our
candidate SN~X has underlying physical parameters \bfPhi$_X$ that are
similar to both \bfPhi$_1$ and \bfPhi$_2$, then we should expect that
the location parameter estimates from those two models should strongly
agree. That is, the location parameter biases from these two models
should be small enough that the true location of our candidate,
\bftheta$_X$, will be simultaneously favored by both $M_1$ {\em and}
$M_2$.  This argues for the use of the intersection operator (Paper~I,
Eq.~17) when combining $g_1$ with $g_2$.

Now suppose that we carry out a third model comparison with model
$M_3$, and let us assume that this model has intrinsically very
different physical parameters \bfPhi$_3$.  The larger separation
between \bfPhi$_3$ and \bfPhi$_X$ means that the light curve shapes
of $M_3$ and SN~X will be substantially different, and therefore $M_3$
will have to introduce a large \bftheta\ bias in order to compensate.
In this case, we should expect that the intersection of 
$g_3=g(\bfD|\theta,M_3)$ with either $g_1$ or $g_2$  will not
provide a useful estimate of the true candidate location. Thus, an
intersection operation would give misleading results, as the outlier
$g_3$ pulls the composite membership distribution away from the real
value of \bftheta$_X$.

If we expect that template mismatches such as $M_3$ are included in
the group of membership functions being combined, then we cannot
justify using the fuzzy AND operator.  If, however, we can exclude
those outliers from our combination group, then the intersection
operation is justified, and should provide a more precise estimate of
\bftheta$_X$.  The $\gamma$ value provides a simple (though somewhat
crude) metric for this purpose. We can exclude templates that have
very different light curve shapes from our candidate by rejecting any
models that yield a $\gamma$ value below some threshold.  For the
validation tests described in the following section we have used a
cutoff of $\gamma < 5\%$.\footnote{The choice of 5\% is somewhat
  arbitrary, though it is connected to the size of the template
  library. With 28 light curves in our library, if they all had equal
  net membership grades then they would all have
  $\gamma_j\sim3.5\%$. Thus, setting the level at 5\% is a moderately
  restrictive choice, and in practice it generally ensures that some 3
  to 10 models are able to pass through the threshold rejection.  If
  the template library had 50 or 100 light curves, then the
  distribution of membership grades would be more diffuse. In that
  case a lower threshold of 4\% or 2\% would be appropriate, in order
  to maintain $\sim$3 to 10 contributing models.}

In Figure~\ref{fig:fuzzyANDdemo}c we show the end result of applying
the fuzzy AND operation to combine membership functions for the SNLS
object 04D2ja.  The resulting composite membership function has a
single peak in the region that all models collectively agree upon as a
likely location for the candidate SN.
In the case of 04D2ja, the parameter estimate from this intersection
of membership functions provides a better estimate than any single
light curve template could provide.  This is true because each
individual template has a location parameter bias that
attempts to compensate for the mis-match in light curve shape. The
length and direction of these bias vectors are distributed more or
less randomly over the space  of \bftheta.  Thus, when many models are
combined together the biases tend to cancel out.  It is important to
note that if we had not used the crude 5\% threshold to cut out severe
mismatches, then a few of the models that are most egregiously biased
would have dominated the intersection operation, pulling the combined
distribution off into a distant corner of \bftheta\ space. In
Section \ref{sec:BiasCorrections} we will discuss a possible method of
calibrating the template library that could alleviate the need for
this threshold cut.

\section{Verification Tests}
\label{sec:VerificationTests}

\subsection{Data Sets}

\begin{deluxetable}{ccl}
  \tablewidth{0pt}
  \tablecaption{Verification Data Subsets}
  \tablecolumns{3}
  \tablehead{ Data set &  $N_{SNe}$ & Description }
  \startdata
    SDSS-A  & 145 & all SNe from \citet{Holtzman:2008} except SN05gj (unfittable) \\
    SDSS-B  & 127 & photometric cut : -10 classified by SOFT as CC;
    -8 with one-sided light curves \\
    SDSS-C  & 116 & photo-cut + spec. cut : -11 with spectroscopic classification ``Ia?''\\[1mm]
    SNLS-A  &  71 & all 71 SNe from \citet{Astier:2006} \\
    SNLS-B  &  57 & spectroscopic cut : -14 with spectroscopic
    classification ``Ia*'' \\[1mm]
    Combo &  198 & photometric cuts : union of SDSS-B with SNLS-A \\
  \enddata
  \label{tab:subsets}
\end{deluxetable}      

To test the accuracy and precision of SOFT estimates for $z$ and \mue,
we used a combination of data from the SDSS-II and SNLS programs. 
We use 146 spectroscopically confirmed Type Ia SNe from
\citet{Holtzman:2008} and  71 from the SNLS project
\citep{Astier:2006}.  The SDSS sample spans a redshift range from
0.01 to 0.45, while the SNLS SNe fall between 0.25 and 1.0. 
From this combined  data set of 217 SNe, we have extracted a
series of subsets that attempt to eliminate objects that are most
problematic for light curve analysis with SOFT. Table~\ref{tab:subsets}
lists the six subsets used.  

The SDSS-A subset rejects only one singularly peculiar object,
SN2005gj.\footnote{This was an extremely luminous SN that appears to
  be a hybrid Type Ia/IIn object \citep{Aldering:2006}.  SOFT is
  unable to match a light curve to this object using any of the Type
  Ia templates it has available, but finds that the best fitting
  template is SN~1994Y, a Type IIn object.}  We define the SDSS-B
sample using photometric selection criteria: excluding ten objects
that are classified as core collapse SNe by SOFT (see Paper I for
details on the classification procedure), and eight objects that have
one-sided light curves, lacking either the pre-peak or the post-peak
epochs entirely.\footnote{ Note that the SDSS-B subset is not truly a
  photometrically selected sample, because spectroscopic criteria were
  used to define the original set of 146 by \citet{Holtzman:2008}.}
The final SDSS subset, SDSS-C, uses an additional cut to remove 11
more objects that were given a spectroscopic classification of ``Ia?''
by \citet{Holtzman:2008}, indicating some classification uncertainty.

The SNLS SNLS-A includes all 71
SNLS SNe presented by \citet{Astier:2006}.  We do not apply a
photometric cut to the SNLS data in the manner of SDSS-B, because SOFT
classifies all 71 SNe as Type Ia, and the peak is well defined for all
light curves.
%\footnote{ The SNLS object 04D2cf is the only one with a
%  light curve that lacks pre-peak data, but it is well fit by SOFT, so
%  there is no need to consider a subset of 70 SNLS SNe.}
For the SNLS-B sample we exclude 14 objects that were
given a ``Ia*'' classification in \citet{Astier:2006} to
indicate some classification ambiguity. 

Our final data set is a combination of the photometrically selected
samples from SDSS and SNLS.  With 127 SNe from SDSS-B and 71 SNe from
SNLS-A, the Combo subset contains 198 objects with complete light
curves that SOFT recognizes as Type Ia.

\subsection{Priors}

For each SN in the verification data sets, we want to use 
Eq.~\ref{eqn:g(D|theta,Mj)} to define fuzzy set membership functions.
To compute the location prior term in that equation,
$p(\bftheta|M_j)$, we use flat uninformative priors
for the parameters \tpk\ and \mue.  For host galaxy extinction we
apply the ``Galactic Line of Sight''  \Av\ prior employed 
by the ESSENCE team \citep{Wood-Vasey:2007}, which is 
based on the host galaxy extinction models of \citet{Hatano:1998}, 
\citet{Commins:2004}, and \citet{Riello:2005} (see Paper I, \S3.1).

For the redshift prior we consider two alternative situations.  
In the first case, we assume a traditional SN survey strategy, where
all SNe have spectroscopic measurements to define their redshifts and
to aid in classification.  The redshift prior for this scenario is
based on the spectroscopic measurements reported by
\citet{Holtzman:2008} and \citet{Astier:2006} for the SDSS and SNLS
data sets, respectively.  The prior is separately defined for each SN
as a Gaussian centered on the measured redshift with width 
defined by $\sigma_z=0.005$.   The parameter estimates that SOFT
returns in this case can be more or less directly compared to the SN
cosmology results elsewhere in the literature. 

Our second redshift prior is designed to evaluate how well SOFT can
measure redshift and distance without any spectroscopic information. 
As a worst case scenario we assume that there is no
spectroscopic information available for any of the SNe or their host
galaxies, and there are not even photometric redshift estimates from
the host galaxies.  
For this case we want to use an uninformative redshift prior,
reflecting our total ignorance of each object's redshift.  One
possibility would be a flat distribution over $z$, normalized so that
it integrates to unity over the range of redshifts under consideration
for each SN candidate.  A slightly more realistic prior would take
into account the increasing volume of space as a function of
redshift.  
%The comoving volume of a shell at redshift z (per unit solid angle) is
%proportional to the square of the comoving distance, times a redshift
%interval dz that defines the shell thickness: 
%
%$$ dV_z \propto  D_C^2   dz $$
%
%\noindent 
%To keep our crude prior simple, we assume a flat
%$\Omega_M=0$ universe,  so that $D_C$ is proportional to z.  In that
%case, our prior for the fraction of all SN explosions that occur
%within the shell at redshift z is: 
For simplicity we assume a a flat $\Omega_M=0$ universe so that the
volume of a shell at redshift z (per unit solid angle) is 

$$ p(z) dz \propto z^2 dz $$

\noindent This is the uninformative prior that we use throughout this
paper, although we also tested the flat z prior and found no
significant changes in our conclusions.  
%A more robust ``worst case''
%prior could be derived by modeling the expected SN yield as a function
%of redshift for SDSS and SNLS, accounting for the depth, cadence,  and
%area of each survey, but this is beyond the scope of the present work.

\subsection{Photometric Redshifts}

\begin{figure}[tb]
  \centering
  \includegraphics[draft=False,width=\columnwidth]{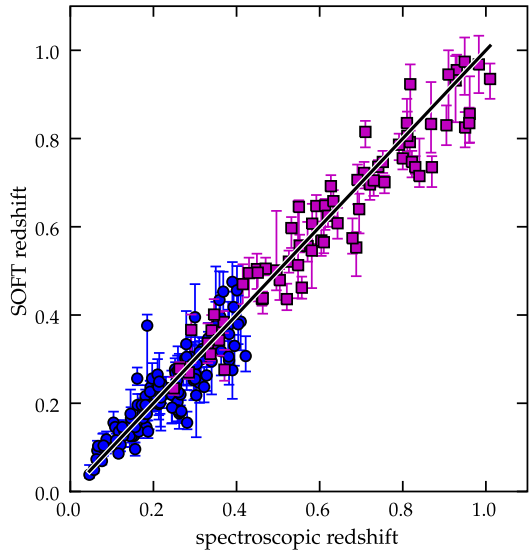} 
  \caption{ 
    Comparison of the peak redshift estimates from SOFT against
    the spectroscopic redshift measurements from SDSS and SNLS.  The
    SDSS SNe are plotted as (blue) squares, and SNLS objects are
    (magenta) circles. The solid line shows how the data would fall
    if SOFT photo-z's traced the spectroscopic redshifts perfectly.
    The RMS scatter about the line is 0.048, and the reduced $\chi^2$  
    statistic is 1.51. 
  }
  \label{fig:combozplot}
\end{figure}

As a first test, we want to evaluate how well SOFT can determine a
redshift based on the SN light curve alone.  For this task we use the
Combined data set, containing 198 photometrically identified Type Ia
SNe from subsets SDSS-B and SNLS-A (see Table~\ref{tab:subsets}).
Using the uninformative $z^2$ redshift prior, we determine a composite
membership function from each \TNSN\ model using Equation
\ref{eqn:g(D|theta,Mj)}.  The membership functions are marginalized
over all nuisance parameters (\Av,\tpk,\mue) and combined using the
fuzzy AND operator to get a composite membership function over
redshift: $g(z)$.  We locate the peak of
this distribution to define $z_{SOFT}$, and assign error bars by
measuring the width of the contour that contains 68\% of the area.

Figure~\ref{fig:combozplot} plots these SOFT redshift values against
the spectroscopically measured redshifts from SDSS and SNLS.   
The residuals defined by $z_{SOFT}-z_{spec}$ have an RMS
scatter about zero of 0.045 for the SDSS-A sample, 0.059 for SNLS-B,
and 0.051 for the combined sample.   
The uncertainties for these data points are decidedly
non-Gaussian and asymmetric.  Nevertheless, the reduced $\chi^2$ 
term gives a rough measure of the accuracy of our uncertainty
estimates, and we find $chi^2/DOF = $ 0.98, 1.25, and 1.08 for the
SDSS-A, SNLS-B, and combined samples, respectively.  These values
being close to unity suggests that the SOFT error bars are accurately
reflecting the true redshift uncertainties.  

An appropriate comparison for these results is the work of
\citet{Oyaizu:2008}, in which photometric redshift errors from
$ugriz$ imaging of $\sim300,000$ SDSS galaxies are evaluated across a
similar range of redshifts. 
Oyaizu et al.\ found that the RMS scatter in photometric
redshift errors ($\delta z=z_{phot}-z_{spec}$) for the set of all
galaxies in their validation set was approximately 0.054.  They also
quantify the photo-z accuracy using $\sigma_{68}$, which measures the
range of the $\delta z$ distribution that contains 68\% of their
validation set objects. The distribution of SDSS galaxy photo-z errors
is more sharply peaked than a Gaussian distribution, but with larger
tails. The $\sigma_{68}$ metric is less sensitive to the fat
tails, so it shows a significantly tighter correlation to the
spectroscopic redshifts, with $\sigma_{68}\approx0.021$.   

From this comparison we see that the galaxy photo-z's and the 
SOFT light curve redshifts can be very complementary techniques for
doing SN science without spectroscopy.   Suppose a SN is detected and
a galaxy is identified as a likely host. We can derive a photo-z from
optical photometry of the host, and also determine a completely
independent photometric redshift using SOFT analysis of the SN light
curve.  In most cases the galaxy photo-z will be more precise (i.e. it
lies in the sharp peak of the photo-z residual distribution).
However, if the photo-z error bar is large, then we should take that
as evidence that the SOFT redshift is more 
reliable (i.e. the host photo-z may be a significant outlier, off in
the fat tails of the residual distribution).   
If the photometric redshift of the presumed host galaxy strongly
disagrees with the SOFT redshift, then we should be suspicious that
the galaxy may be a chance superposition and may not actually be the
SN's host.  Finally, in cases where the host galaxy is too faint for
detection, the SOFT light curve redshift can serve as a reliable
stand-in. 

In practice, host galaxy photo-z's can be fed into the SOFT program as
a redshift prior, in exactly the same way as spectroscopic redshifts.
SOFT will naturally balance that host galaxy prior against the
posterior information provided by the SN light curve -- just as it is
done in purely Bayesian approaches.  When using priors drawn from host
galaxy redshifts (be they photometric or spectroscopic), some care
must be taken to avoid introducing a redshift bias.  If the redshift 
distribution of the galaxies is significantly different from the
redshift distribution of the SN sample being studied, then the
likelihood of chance superpositions is increased. For example, the
photo-z catalog of \citet{Oyaizu:2008} is primarily comprised of
$z<0.4$ galaxies, so if that catalog were used to provide redshift
priors for a survey like the SNLS then it would lead to underestimated
redshifts for all the high-z SNe.
 
\subsection{Photometric Cosmology}

In addition to the photometric redshift verification test, we have
also evaluated the ability of SOFT to provide simultaneous estimates
of redshift and distance, for testing cosmological models. 
From each SDSS and SNLS light curve we derived a composite
membership function in the (z,\mue) plane, in the manner depicted in
Figure~\ref{fig:fuzzyANDdemo}.  From the resulting composite
membership function, we extracted an estimate for z and
\mue\ by locating the peak of the composite membership function.  
To estimate the uncertainty we measured the height and width of the
``1-$\sigma$ contour,'' which we define as the border of the smallest
contiguous region about the peak that contains 68\% of the integrated
value of the composite membership function. 
Figure~\ref{fig:fuzzyANDdemo}c shows an example of the 1-$\sigma$
contour from a composite membership function. 

\begin{figure}[!tb]
  \centering
  \includegraphics[draft=False,width=\columnwidth]{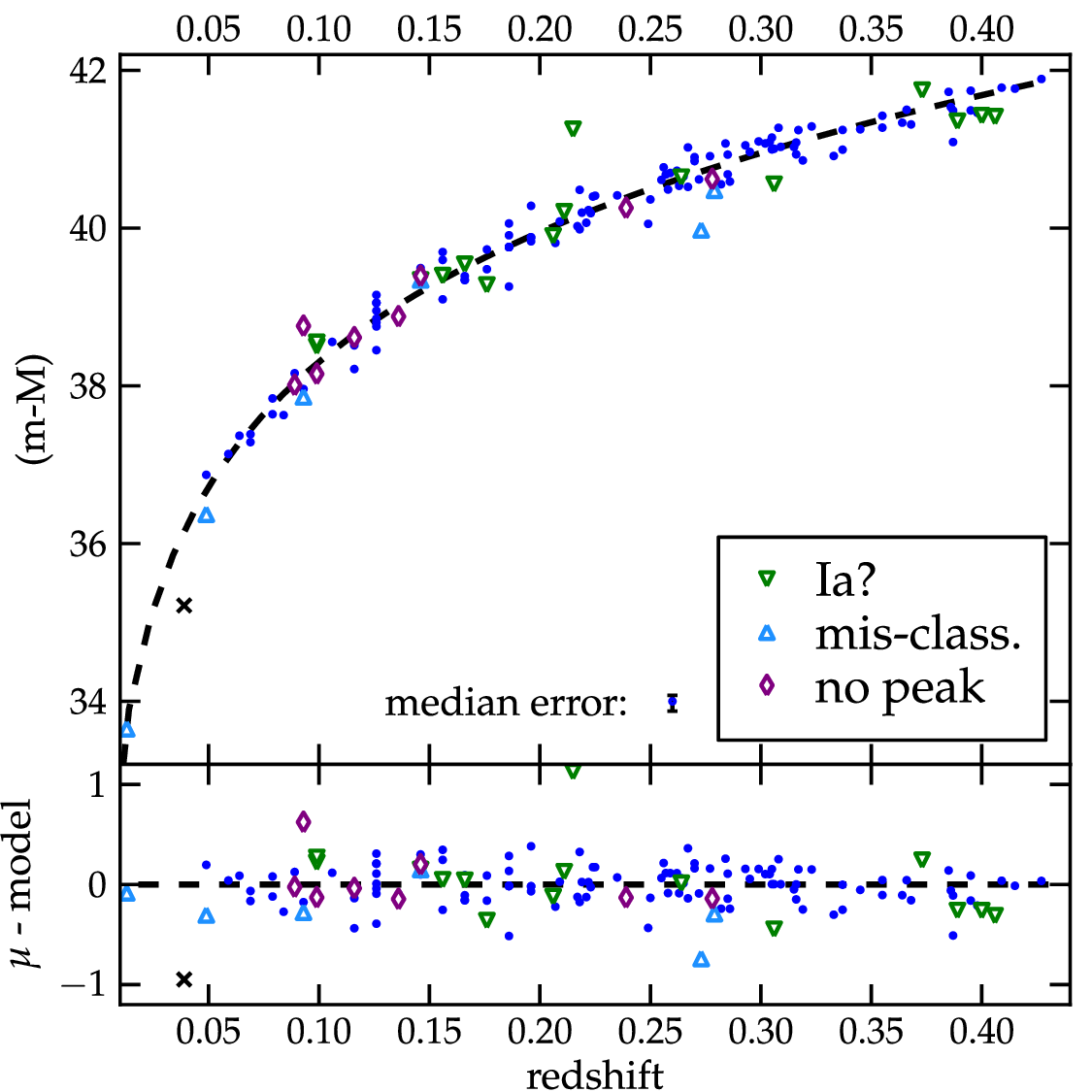} 
  \caption{ Hubble diagram showing SOFT parameter estimates from the
    SDSS verification data set, using spectroscopic redshift priors.  
    All 146 of the SDSS-II SNe survey are plotted,
    with different symbols to indicate the classification cuts that
    define SDSS subsets A, B and C.  Median error bars for
    \mue\ are shown in the lower right.
    The 16 SNe with ambiguous
    spectroscopic classifications (the ``Type Ia?'' objects) are
    plotted as (blue) squares. The (cyan) triangles indicated 10 objects
    that SOFT classifies as Core Collapse SNe (5 of these are also in
    the ``Ia?'' set).   An additional 8
    objects that have one-sided light curves are plotted as (purple)
    diamonds.  The X symbol marks SN\,2005gj, a superluminous ``Type
    IIa'' object, which SOFT is unable to match with any Type Ia light
    curve.  It is plotted at the position that SOFT would assign based
    on the best-fit template, which is the Type IIn SN\,1994Y.
    In the lower panel we show the residual distance modulus
    values as a function of redshift. Goodness of fit statistics for
    the comparison of these data against the $\Lambda$CDM reference
    model are provided in Table~\ref{tab:fitstats}.
  }
  \label{fig:sdssHubble}
\end{figure}

\begin{figure}[tb]
  \centering
  \includegraphics[draft=False,width=\columnwidth]{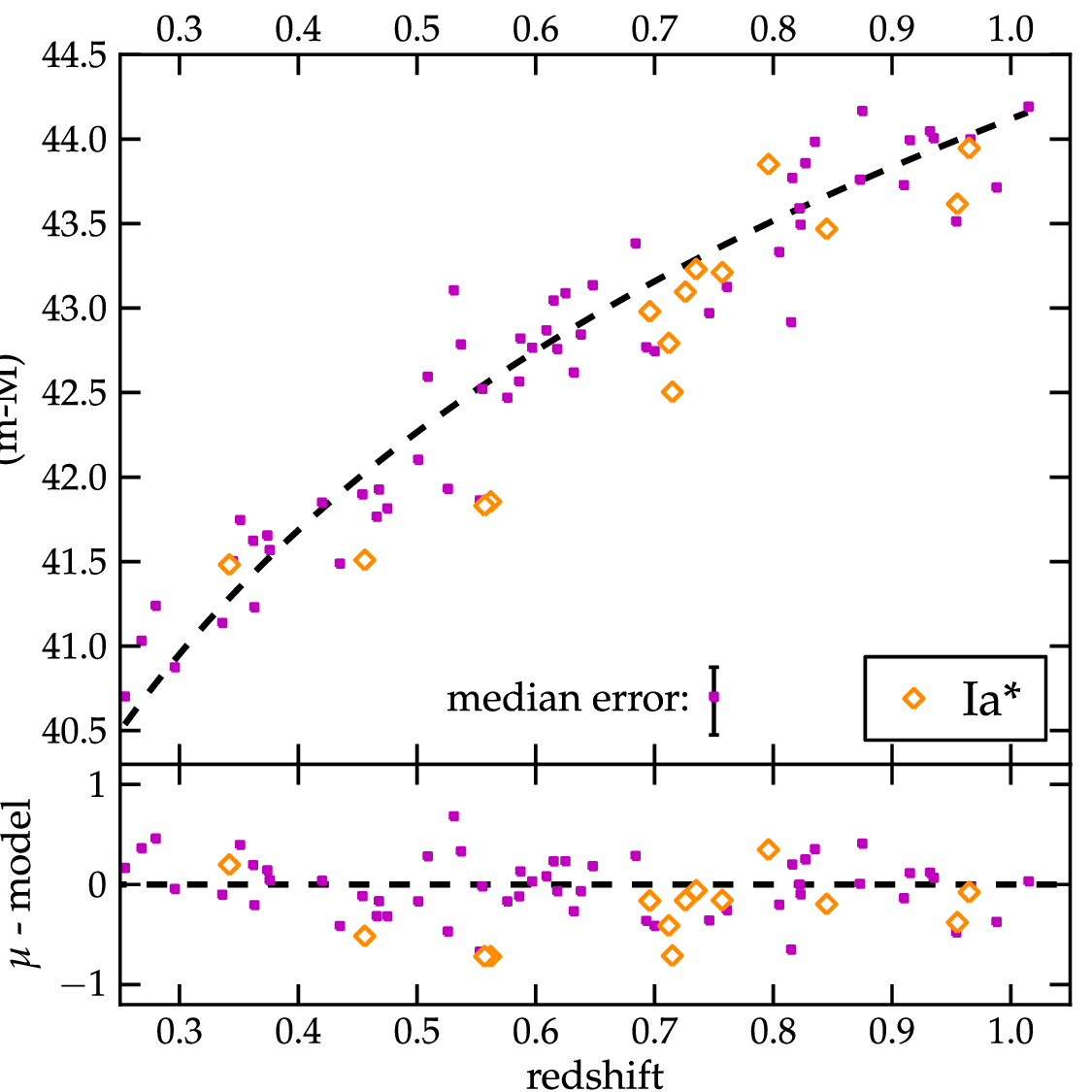} 
  \caption{ Hubble diagram showing SOFT parameter estimates of the
    SNLS verification data set, as in Fig.~\ref{fig:sdssHubble}. 
    For 14 objects with uncertain spectroscopic classifications (the
    ``Type Ia*'' SNe) we plot (magenta) squares.  A median error bar
    is shown in the lower right, and residuals are plotted in the
    lower panel.  Goodness of fit statistics from
    the comparison of these data against the $\Lambda$CDM reference
    model are given in Table~\ref{tab:fitstats}.
  }
  \label{fig:snlsHubble}
\end{figure}

In Figure~\ref{fig:sdssHubble} we show a Hubble diagram constructed
with SOFT parameter estimates from all 146 SDSS SNe, using
spectroscopic redshift priors. As described in the caption, different
symbols indicate the objects that are excluded by photometric and
spectroscopic cuts when culling down to the SDSS-A, B, and C subsets.  
Error bars have been withheld for clarity, so a representative error
bar is shown in the lower right. 
Overlaid on the data points is a dashed line
representing the $\Lambda$CDM cosmological model that is currently
favored by observational data, with 
$H_0=70.5\,km\,s^{-1}$, $\Omega_M=0.273$, and $\Omega_\Lambda=0.726$
\citep{Komatsu:2009}.   Note that this model is presented here only as
a metric for evaluating the SOFT results, and it was not generated by
fitting to these data. Although the SOFT parameter estimates could be 
used to determine the best-fitting cosmological model for these
particular data, such analysis is beyond the scope of this paper. 
The Hubble diagram in Figure~\ref{fig:snlsHubble} shows the SOFT
results from all 71  SNLS SNe, also computed using spectroscopic
redshift priors. 

\begin{figure}[tb]
  \centering
  \includegraphics[draft=False,width=\columnwidth]{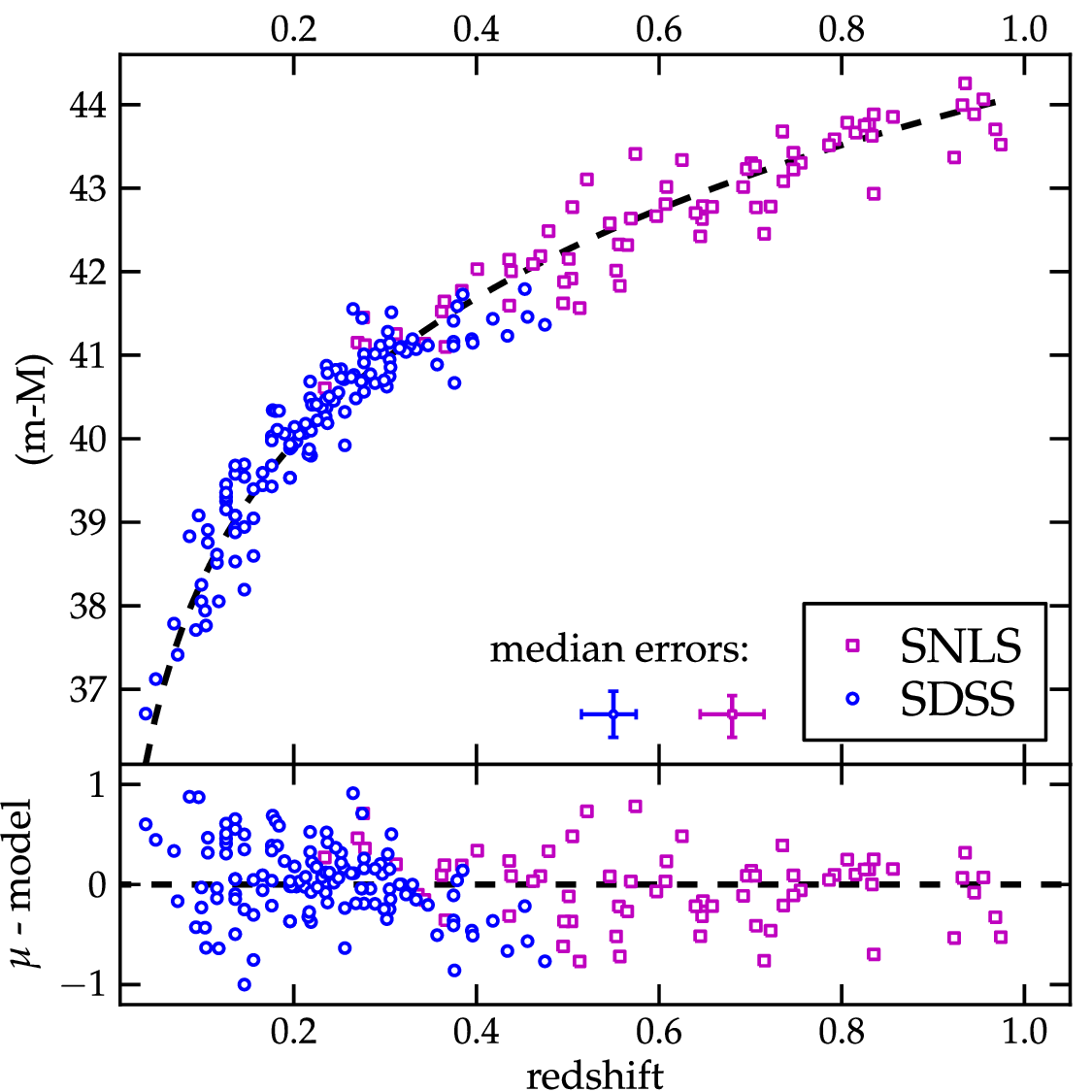} 
  \caption{ Hubble diagram from the Combined verification data set
    (see Table~\ref{tab:subsets}).  All 
    redshifts and distances are derived by SOFT alone, using the
    uninformative $z^2$ redshift prior.  Objects with one-sided light
    curves and those which SOFT classifies
    as core collapse SNe have been rejected, but no spectroscopic
    classification cuts were used.  Goodness of fit statistics from a
    comparison against the $\Lambda$CDM reference model are shown in
    Table~\ref{tab:fitstats}. 
  }
  \label{fig:comboHubble}
\end{figure}

To quantify the comparison of our SOFT parameter estimates against the
\citealt{Komatsu:2009} model, we compute the  $\chi^2$ goodness-of-fit
statistic. For the distance modulus uncertainty
we use the quadratic sum of the positive and negative
\mue\ uncertainties and a redshift uncertainty component: 

\begin{equation}
  \sigma_\mu^2 = \sigma_{\mu,{\rm pos}}^2  + \sigma_{\mu,{\rm neg}}^2  + \sigma_{\mu,z}^2 
\label{eqn:sigmamu}
\end{equation}

\noindent For simplicity, we have projected the redshift
uncertainty of each object onto the \mue\ axis using

\begin{equation}
 \sigma_{\mu,z}^2 = \sigma_z^2 \left(\frac{5}{ln(10)}\right)\frac{(1+z)}{z(1+z/2)} 
\label{eqn:sigmamuz}
\end{equation}

\noindent in which  $\sigma_z^2 = \sigma_{z,{\rm pos}}^2 +
\sigma_{z,{\rm neg}}^2$ is
the quadratic sum of positive and negative redshift errors derived
from the 1$\sigma$ contours. 

In addition to the reduced $\chi^2$ test, we measure the scatter about
the model by computing the root mean square Hubble residual values:

\begin{equation}
RMS_\mu = \sqrt{ \frac{\sum\limits_i (\mu_e^{SOFT} - \mu_e^{model})^2}{N}}
\label{eqn:RMSmu}
\end{equation}

\noindent where $N$ is the number of SNe in the sample.   A summary of
these fit quality statistics for all data subsets is provided in
Table~\ref{tab:fitstats}. 

\begin{deluxetable}{ccccc}
  \tablewidth{0pt}
  \tablecaption{Hubble Diagram : Goodness of Fit Statistics}
  \tablecolumns{5}
  \tablehead{ data set &  $\chi_\mu^2$  & $N_{DOF}$ & $\chi^2/N$ & RMS$_\mu$}    
  \startdata
    \multicolumn{5}{c}{\bf with spectroscopic priors} \\[1mm]
    SDSS-A  & 504.5 & 142 & 3.55 & 0.23 \\
    SDSS-B  & 398.3 & 124 & 3.21 & 0.21 \\ 
    SDSS-C  & 178.1 & 113 & 1.58 & 0.18 \\[1mm]
    SNLS-A  & 135.8 & 68  & 1.99 & 0.31 \\
    SNLS-B  &  93.5 & 54  & 1.73 & 0.28 \\
    \tableline
    \multicolumn{5}{c}{\bf no spectroscopic information} \\[1mm]
    SDSS-A  & 272.8 & 142 & 1.92 & 0.43 \\
    SDSS-B  & 141.1 & 124 & 1.14 & 0.38 \\[1mm]
    SNLS-A  &  96.4 & 68  & 1.42 & 0.35 \\[1mm]
    Combined   & 238.5 & 195 & 1.22 & 0.37 \\
  \enddata
  \label{tab:fitstats}
\end{deluxetable}

We can see from Figures~\ref{fig:sdssHubble} and \ref{fig:snlsHubble}
and from Table~\ref{tab:fitstats} that the photometric and
spectroscopic cuts are doing a good job of removing serious
outliers: the $\chi^2$ and $RMS$ statistics improve each time a set of
suspect SNe is removed. 
When spectroscopic redshift priors are included, the RMS
scatter from SOFT approaches the precision that can be achieved with
other light curve fitters.  As a recent example,
\citet{Kessler:2009} applied MLCS2k2 and SALT-II to approximately the
same data. Using 103 objects from the SDSS sample (comparable to our
SDSS-C subset), Kessler et al.\ found  $RMS_\mu=0.15$ mags when using
MLCS2k2.  With 62 SNe from the SNLS sample (similar to our SNLS-B),
they get $RMS_\mu=0.24$ mags.\footnote{See Table 11 of
  \citet{Kessler:2009}, but note that the $\chi^2_\mu$ values that
  they report are calculated with an additional internal error of
  $\sigma_{int}^\mu=0.16$, which we have not duplicated.}

As is to be expected, when SOFT has no spectroscopic information to
use for weeding out misclassifications and for defining redshift
priors, the resulting fit statistics are degraded.  However, even with
no spectroscopic help, SOFT is still able to produce consistent and
reliable redshift and distance estimates across the entire range of
z. The spectroscopy-free Hubble diagram of
Figure~\ref{fig:comboHubble} shows that the increased RMS scatter is
not being driven by large systematic shifts.  The fact that reduced
$\chi^2$ statistics in the lower half of Table~\ref{tab:fitstats} are
close to unity indicates that the SOFT uncertainty estimates are doing
a good job of representing the true error in the z and \mue\ parameter
estimates.  Recall that our redshift priors for this spectroscopy-free
cosmology test were constructed under the assumption of a ``worst-case
scenario,'' with no host galaxy photo-z estimates at all.  In reality,
many SNe from future wide-field surveys will have good photometric
redshifts from their host galaxies, which can significantly improve
the SOFT parameter estimates. Those future wide-field surveys will
also have much larger data samples, helping to reduce the parameter
estimate uncertainties that arise from random effects.  As the sample
size increases, however, systematic uncertainties will begin to
dominate.  In Section~\ref{sec:BiasCorrections} we will consider
potential modifications of the SOFT method that may help to minimize
systematic errors.

%%% INSERT MONTE CARLO BOOTSTRAP SECTION

\section{Variable Dark Energy}
\label{sec:VariableDarkEnergy}

One of the primary reasons for developing a tool such as SOFT is to
deal with the extremely large SN data sets from a new generation of
synoptic surveys.  A principal science goal for these next generation
surveys is to improve constraints on a time-variable dark energy (DE)
equation of state.  In this section we ask: How much
of an improvement can be made with the precision provided by SOFT?

As a foundation for our simulations we must first adopt a DE
model in which the equation of state parameter, $w$, is a function of
the scale factor $a$, and therefore varies over cosmic time.
A simple and commonly used parameterization for $w(a)$ is to assume a
linear form \citep{Chevallier:2001,Linder:2003}:

\begin{equation}
   w(a) = w_0 + w_a(1-a)
\label{eqn:w(a)}
\end{equation}

\noindent For our simulations we add the additional restriction of a
flat universe (\Ok=0). The Friedmann equation in this case is: 

\begin{equation}
  \begin{array}{ll}
   \frac{H(z)}{H_0} = [&\Om (1+z)^3 + \\
      &  (1-\Om)~{(1+z)^{3(1+w_0+w_a)}}~{e^{-3 w_a z}} ]^{1/2} \\
  \end{array}
   \label{eqn:FwaCDM}
\end{equation}

\subsection{Bootstrap Monte Carlo}
\label{sec:BootstrapMonteCarlo}

To investigate the informative power of future SN data sets, we use a
series of Monte Carlo simulations in which the real SOFT results from
our validation tests serve as seed points for creating larger
simulated data sets.  Our goal is to mimic the observational data set
that might be produced by PS1, DES, LSST, or another large and deep SN
survey.  The SDSS and SNLS data provide a good baseline for simulating
those future surveys because they have similar light curve sampling
and cover a similar redshift range.

We create nine Monte Carlo data sets, labeled A through I, using the
time-varying DE model with cosmological parameters fixed at
$(\Om,\Ode,w_0,w_a) = (0.3,0.7,-0.5,-1.0)$.
The nine data sets are distinguished according to the
number of SNe in the sample, and the source of redshift information
for each object.  The A, B and C data sets have 500 SNe; D, E, and F
have 2000; while G, H, and I have 10000 objects.  The first column
(A,D,G) assumes an uninformative $z^2$ prior for the redshift. The
second column (B,E,H) uses (simulated) host-galaxy photometric
redshifts.  The final column (C,F,I) uses spectroscopic redshift
priors. 

%Each point in one of our Monte Carlo simulations is a clone of a real
%SN observation, drawn at random from the combined SDSS$+$SNLS data
%set. In order to project these SDSS and SNLS objects onto an arbitrary
%cosmology, we must first assign each one a ``true'' redshift and
%distance.  For redshift we can use the directly measured spectroscopic
%$z$ values. To translate that $z$ into a true distance parameter \mue,
%we adopt the cosmological model of \citet{Komatsu:2009}, which
%we have already seen is a good fit for these data.

The process of ``bootstrapping'' a cloned observation out of the
seed data is as follows:

\begin{enumerate}
\item Draw a ``seed'' SN at random from the SDSS$+$SNLS data.
\item Assume that the spectroscopic redshift, $z_{\rm SPEC}$, is equal
  to the ``true'' redshift, $z_{\rm TRUE}$, and use cosmological
  parameters from the \citet{Komatsu:2009} model to determine the
  ``true'' distance modulus $\mu_{\rm TRUE}$.
\item Collect the SOFT results ($z_{\rm SOFT}$,$\mu_{\rm SOFT}$) for
  the seed object, using the appropriate redshift prior ($z^2$, host
  photo-$z$, spec-$z$).
\item Determine the error on the redshift and distance:
$z_{\rm err}=z_{\rm SOFT} - z_{\rm TRUE}$  and $\mu_{\rm err}=\mu_{\rm SOFT} - \mu_{\rm TRUE}$ .
\item Generate a clone point at a similar redshift 
$z_{\rm CLONE} = z_{\rm TRUE} + z_{\rm OFFSET}$, where 
$z_{\rm OFFSET}$ is drawn from a normal distribution with 
standard deviation $\sigma=0.05$.
\item Determine the clone's true distance modulus $\mu_{\rm CLONE}$
  based on its redshift, using the variable DE model with $w_0=-0.5$
  and $w_a=-1.0$.
\item Add an offset to $z$ and $\mu$ to account for observational
  error. These offsets are drawn from normal distributions centered on
  the SOFT error values $z_{\rm err}$ and $\mu_{\rm err}$ from step 4,
  with the distribution width reflecting the SOFT uncertainty values:
  $\sigma_z=\delta z_{\rm SOFT}$ and $\sigma_\mu=\delta \mu_{\rm
    SOFT}$.
\item Generate an error bar for the clone point in both the $z$ and
  the $\mue$ directions using the SOFT errors from the seed object,
  augmented by a random adjustment drawn from normal distributions
  with $\sigma_z=\epsilon \cdot \delta z_{\rm SOFT}$ and
  $\sigma_\mu=\epsilon \cdot \delta \mu_{\rm SOFT}$, where $\epsilon$ =
  0.05, 0.1, and 0.2 for the $z^2$, photo-z, and spec-z priors,
  respectively.
\end{enumerate}

At the end of this process we have a new cloned data point, which has
been ``observed'' at a position $z_{\rm obs},\mu_{\rm obs}$.  This
redshift and distance pair (along with associated uncertainties)
incorporates the desired variable DE model, and also reflects the real
errors and observational uncertainties of the original seed object.
The sequence is repeated N times (where N=500, 2000, or 10000) to
generate a complete Monte Carlo sample.  With each bootstrap Monte
Carlo data set we then fit for cosmological parameters using the
variable DE model.  

In addition to the simulated SN data, these
cosmological fits are constrained by observations of Baryon Acoustic
Oscillations (BAO) and the Cosmic Microwave Background (CMB). 
We use the BAO constraint on the sound barrier $r_s/D_V$ from
SDSS DR7 \citep{Percival:2010} and the CMB constraint on the ``shift
parameter'' $R$ from the 7-year WMAP results \citep{Komatsu:2010}.

\subsection{Results}
\label{sec:VarDEResults}

\begin{figure*}[p]
  \centering
  \includegraphics[draft=False,width=\textwidth]{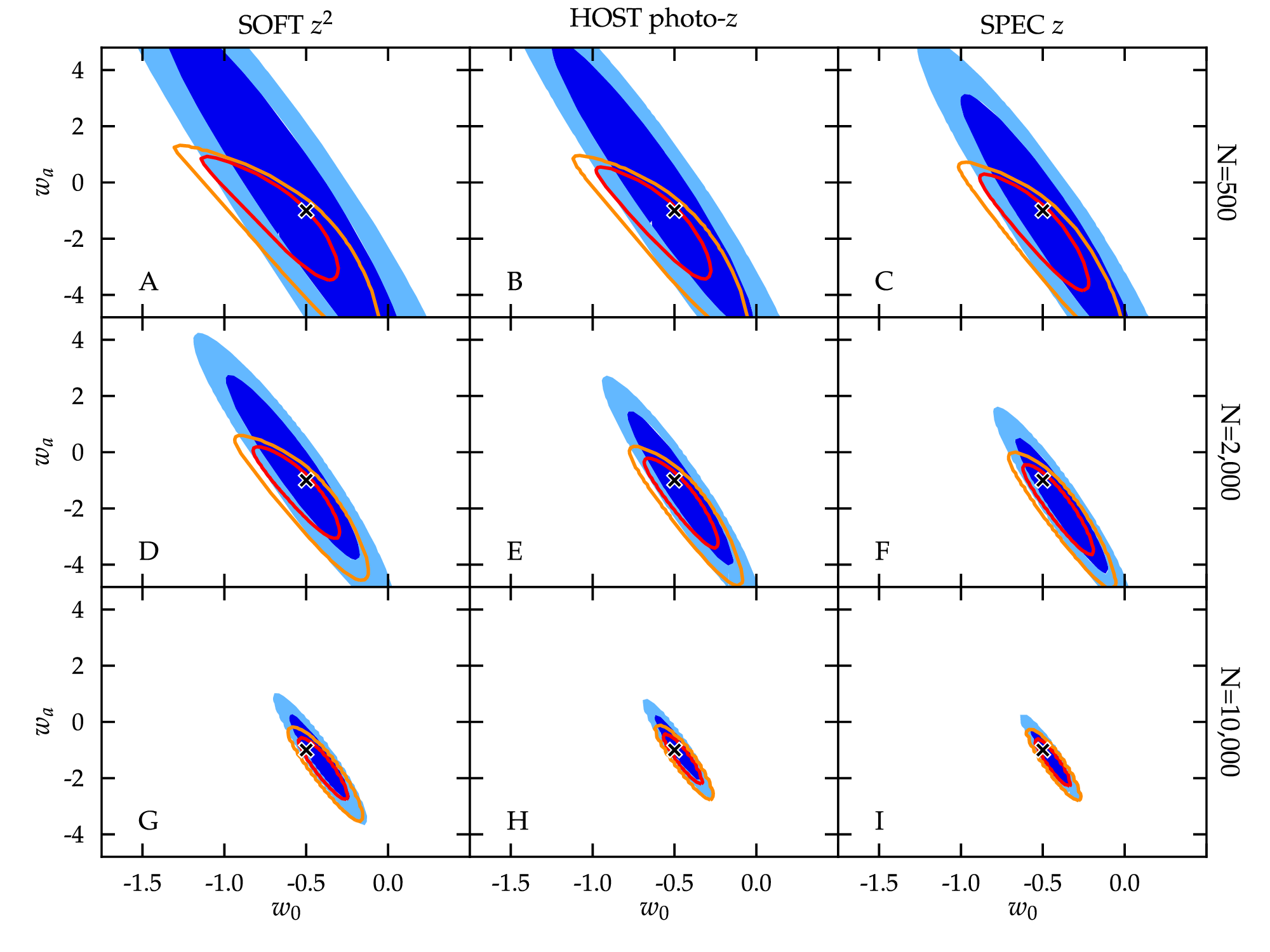} 
  \caption[Likelihood contours from the dark energy Monte Carlo
    simulations]{ Likelihood contours in ($w_0$,$w_a$) space for the
    dark energy ``Bootstrap Monte Carlo'' simulations, based on data
    from SDSS and SNLS. Filled contours (in blue) show the 95\% and
    99.9\% confidence boundaries from the SN data alone. Unfilled
    contours (in red and orange) show the combined constraint from SN,
    CMB, and BAO observations.  A cross in each panel marks
    the location of the input cosmological parameters $w_0=-0.5$ and
    $w_a=-1.0$, which were used to generate the simulated
    observations.  Each row (ABC, DEF and GHI) shows a set of
    simulations with 500, 2000, and 10000 SNe, respectively.  The
    columns show results from using three different sources for the
    redshift priors. The first column (ADG) uses only the very weak
    $z^2$ prior.  In the second column (BEH) we assume every simulated
    SN has a photometric redshift estimate from its host galaxy, and
    in the third column (CFI) we apply a strong spectroscopic prior.
  }
  \label{fig:darkmcContours}
\end{figure*}

The likelihood contours from all nine Monte Carlo simulations are
shown in Figure~\ref{fig:darkmcContours} and the maximum likelihood
values are summarized in Table~\ref{tab:darkmcResults}.  The error
contours and quoted uncertainties reflect statistical errors only.  As
is to be expected, the precision of the DE constraints improves as the
sample size is increased, and as the redshift prior is strengthened.
A convenient quantity for comparing the DE constraints from different
experiments or simulations is the Figure of Merit (FoM) proposed by
the Dark Energy Task Force (DETF).  The DETF FoM is defined as the
reciprocal of the area in the $w_0-w_a$ plane that encloses the 95\%
confidence limit region \citep{Albrecht:2006}.  In
Table~\ref{tab:darkmcResults} we report the FoM values achieved with
from the SN data alone, as well as the FoM from combined SN+CMB+BAO
constraints.  Both values are normalized to the corresponding FoM from
simulation C.  This simulation, with spectroscopic constraints from 500
SNe, roughly corresponds to the current state of the art
\citep[cf][]{Davis:2007,Sollerman:2009,Komatsu:2009,Komatsu:2010}.

\begin{deluxetable}{l|cc@{\hspace{8mm}}cc@{\hspace{8mm}}cc}
  \tablewidth{0pt}
  \tabletypesize{\small}
  \tablecaption{Dark Energy Monte Carlo Simulation Results}
  \tablecolumns{7}
  \tablehead{ 
    \multirow{2}{*}{$N_{\rm SN}$} & \multicolumn{6}{c}{Redshift prior source} \\
       & \multicolumn{2}{c}{SOFT $z^2$} & \multicolumn{2}{c}{HOST photo-$z$} & \multicolumn{2}{c}{SPEC-$z$}%\\[-6mm]
    %\multicolumn{1}{c}{$N_{\rm SN}$} & \multicolumn{2}{c}{SOFT $z^2$} & \multicolumn{2}{c}{HOST photo-$z$} & \multicolumn{2}{c}{SPEC-$z$}
  }

  \startdata
  \multirow{2}{*}{500}       & \multirow{4}{*}{A} & $w_0 =  -0.67 ~^{+0.35}_{-0.48}$ & \multirow{4}{*}{B} & $w_0 =  -0.61 ~^{+0.31}_{-0.38}$ & \multirow{4}{*}{C} & $w_0 =  -0.54 ~^{+0.29}_{-0.34}$ \\[2mm]
                             &                    & $wa =  -0.86 ~^{+1.66}_{-2.64}$  &                    & $w_a =  -1.02 ~^{+1.52}_{-2.38}$ &                    & $w_a =  -1.36 ~^{+1.56}_{-2.54}$ \\[2mm]
                             &                    & FoM$_{\rm SN}$ = 0.59           &                    & FoM$_{\rm SN}$  = 0.82         &                    & FoM$_{\rm SN}$  = 1.0           \\[1mm]  
                             &                    & FoM$_{\rm ALL}$ = 0.81          &                    & FoM$_{\rm ALL}$ = 0.96          &                   & FoM$_{\rm ALL}$ = 1.0            \\[4mm]  
  \multirow{2}{*}{2,000}     & \multirow{4}{*}{D} & $w_0 =  -0.57 ~^{+0.25}_{-0.25}$ & \multirow{4}{*}{E} & $w_0 =  -0.48 ~^{+0.21}_{-0.20}$ & \multirow{4}{*}{F} & $w_0 =  -0.42 ~^{+0.21}_{-0.20}$ \\[2mm]
                             &                    & $w_a =  -1.06 ~^{+1.16}_{-2.04}$ &                    & $w_a =  -1.48 ~^{+1.18}_{-1.92}$ &                    & $w_a =  -1.82 ~^{+1.32}_{-1.78}$ \\[2mm]
                             &                    & FoM$_{\rm SN}$ = 1.75           &                    & FoM$_{\rm SN}$  = 2.74         &                    & FoM$_{\rm SN}$  = 3.62           \\[1mm]  
                             &                    & FoM$_{\rm ALL}$ = 1.56          &                    & FoM$_{\rm ALL}$ = 1.97          &                   & FoM$_{\rm ALL}$ = 2.21            \\[4mm]  
  \multirow{2}{*}{10,000}    & \multirow{4}{*}{G} & $w_0 =  -0.42 ~^{+0.15}_{-0.12}$ & \multirow{4}{*}{H} & $w_0 =  -0.48 ~^{+0.13}_{-0.09}$ & \multirow{4}{*}{I} & $w_0 =  -0.46 ~^{+0.11}_{-0.08}$ \\[2mm]
                             &                    & $w_a =  -1.48 ~^{+0.78}_{-1.32}$ &                    & $w_a =  -1.17 ~^{+0.67}_{-1.03}$ &                    & $w_a =  -1.36 ~^{+0.66}_{-0.94}$ \\[2mm]
                             &                    & FoM$_{\rm SN}$ = 8.21           &                    & FoM$_{\rm SN}$  = 14.47        &                    & FoM$_{\rm SN}$  = 18.60          \\[1mm]  
                             &                    & FoM$_{\rm ALL}$ = 4.51          &                    & FoM$_{\rm ALL}$ = 7.38          &                   & FoM$_{\rm ALL}$ = 8.67            \\[1mm]  
  \enddata

  \tablenotetext{*}{ Cell values indicate the label of the simulation
    set (A--I), the maximum likelihood values for the DE parameters
    $w_0$ and $w_a$, and the DETF Figure of Merit (FoM). 
    The FoM$_{\rm SN}$ value is derived from only the SN constraints,
    while the FoM$_{\rm ALL}$ uses the combined SN+CMB+BAO
    constraints. Both values are normalized to the results from
    simulation C, which approximates the current SN sample.
    Figure~\ref{fig:darkmcContours} for images of the confidence limit
    contours.  }
  \label{tab:darkmcResults}
\end{deluxetable}

%To put the results of these DE simulations into context, we should
%consider how the sample sizes and spectroscopic resources compare to
%present and future surveys.  
%Several recently completed intermediate redshift surveys (e.g. SNLS
%and ESSENCE) will increase the number of spectroscopically observed
%SNe with complete multi-color light curves to $\sim 500$, even after
%applying the most stringent selection cuts.  

The new up-and-coming surveys such as PS1, the Palomar Transient
Factory, and Skymapper are best represented by simulations D
and E, for which the FoM is improved by about a factor of 2. Not an
insignificant gain, to be sure, but one that could easily be washed
out by systematic uncertainties if they are not carefully controlled. 
The next leap forward, to be realized in the next decade by
Pan-STARRS 4, JDEM and LSST, will bring the SN sample size to 10,000
or more objects, as in simulations G and H.

Using the DETF FoM as our principal barometer, our simulations suggest
that a spectroscopy-free analysis using SOFT should be sufficient to
capitalize on the very large SN samples coming in the next decade.
With each new generation of SN surveys the sample size increases by a
factor of 4--5, and the SOFT-based FoM increases by a factor of
2--3. A close look at the uncertainties on $w_0$ and $w_a$ in
Table~\ref{tab:darkmcResults} highlights the fact that systematic
effects will quickly become a dominant source of uncertainty.  For a
sample size of 2,000 objects the statistical uncertainty on $w_0$ from
our Monte Carlo simulations is around $\pm 0.2$, which is already
comparable to the expected level of systematic uncertainty.

How can these results inform our strategic choices for new and
upcoming SN cosmology projects?  We would argue that the new surveys
can be successful relying on light-curve-based redshifts derived by
SOFT or similar programs, informed by host galaxy photo-$z$'s when
possible.  For spectroscopic follow-up, instead of targeting the most
likely \TNSNe, cosmological analyses will be better served by focusing
resources on objects that are given an ambiguous classification grade
by SOFT.  Most of these objects will be SNe of type Ib/c or Ia,
perhaps with unusual colors or anomalous luminosities that make them
difficult to classify.  To the extent that spectroscopic measurements
are able to provide a more definitive classification, we can sharpen
SOFT's ability to discriminate such marginal cases and help to reduce
systematic errors from misclassifications.

In spite of the close connection to real data, the results of our
bootstrap Monte Carlo approach are likely to be overly optimistic.  We
are necessarily overlooking the effects of sample selection and SN
classification, and we have knowingly omitted systematic uncertainties
from our analysis.  In order to realize these optimistic predictions,
it will be necessary to make adjustments and improvements to the SOFT
program itself that will improve its precision and reduce systematic
effects.  In the following section we provide an outline for that
future development.

%Although these Monte Carlo simulations are optimistic on their face,
%they also make no attempt to predict any possible improvement in the
%SOFT method itself. Most importantly, we are using estimates for the
%SOFT precision that are based on the use of $\sim$25 \TNSN\ templates
%(see Paper I). This template library will certainly be expanded in
%coming years, and we can expect a subsequent improvement in the
%precision that can be achieved with SOFT.

%Perhaps more important than an abundance of spectroscopic observations
%will be a substantial contribution of infrared imaging.  Extinction
%from interstellar (or circumstellar) dust is presently the most
%important systematic effect for SOFT, and that uncertainty could be
%reduced by setting well-informed priors for $R_V$ and $A_V$.
%Measurements of these extinction parameters are much improved by
%including photometry of the rest-frame near-infrared (NIR) along with
%the usual rest-frame optical data \citep[e.g.][]{Krisciunas:2007}.

%%% END BOOTSTRAP MONTE CARLO

\section{Bias Corrections}
\label{sec:BiasCorrections}

As discussed in Section \ref{sec:ParameterEstimation}, our use of
fixed shape light curve templates introduces an inherent bias into the
parameter estimates that can be derived from each individual template.
In this work we have addressed this problem by using the fuzzy
intersection operator when combining membership functions.  The fuzzy
AND emphasizes collective agreement from all contributing templates,
and therefore generally causes the individual biases to cancel out.
In this section we consider two separate approaches that might reduce
the effects of individual template biases by detecting and
pre-emptively correcting for those location parameter offsets.

\subsection{Template Calibration}
\label{sec:Template Calibration}

\begin{figure}[tb]
  \centering
  \includegraphics[draft=False,width=\columnwidth]{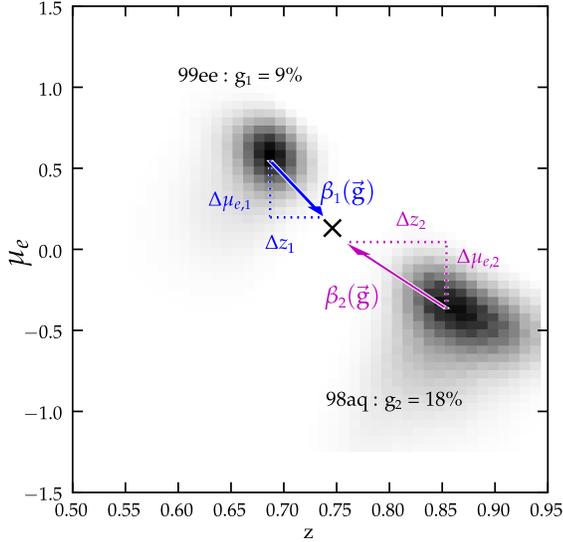} 
  \caption{ Illustration of the bias correction vectors $\beta_j$ that
    could be derived from a calibration of the SOFT template library.
    The background image shows the fuzzy membership functions for SNLS
    04D2ja (as seen in Figure~\ref{fig:fuzzyANDdemo}) from two
    separate template comparisons: SN 1999ee and 1998aq. The
    integrated membership grade $\gamma$ for each template is listed
    as a percentage. The goal of a template calibration process would
    be to derive a separate bias correction vector $\beta_j$ for each
    template. For any given candidate the size and direction of each
    template's bias correction would be based on the unique pattern of
    integrated membership grades: ${\bf \Gamma}=\{\gamma_1, \gamma_2,
    \ldots \gamma_N\}$.  }
  \label{fig:biasDemo}
\end{figure}

Figure~\ref{fig:biasDemo} illustrates how one could use a SN with a
known location to measure the size and direction of template biases.
The redshift and distance for our supposed calibration SN is marked by
the ``x.''  After a SOFT comparison of
our calibration SN against the model $M_1$ (based on SN 1999ee in
Figure~\ref{fig:biasDemo}), we can measure the peak
location of the $M_1$ cloud as $(z_1,\mu_{e,1})$.  Knowing the true
location of the SN, we can then measure the distance from the cloud to
the true position of the SN as $(\Delta z_1,\Delta \mu_{e,1})$.  This
immediately gives us a bias correction vector $\beta_1 = (-\Delta
z_1,-\Delta \mu_{e,1})$ that can be used to translate the entire
probability cloud down to the true location of the peak.  We can
compute similar correction vectors for model $M_2$ (as shown in
Figure~\ref{fig:biasDemo}) and all other models. 
%giving us a complete set of correction terms that will separately
%improve the accuracy of each individual template's parameter estimate.

Unfortunately, once we derive a set of bias correction vectors using a
particular calibration object, we can only apply them to other
candidate SNe that are basically identical to the calibration
object. If we have a candidate SN that is physically more similar to
the model $M_1$ than our calibrator, then the $\beta_1$ vector shown
in Figure~\ref{fig:biasDemo} would be an over-correction. To avoid
excessive bias correction, we need to scale the $\beta$
values separately for each candidate SN according to its similarity
with the calibration object.  
For this purpose, we can use the set of all integrated membership
function values from Eq.~\ref{eqn:gammaj}:
${\bf \Gamma}=\{\gamma_1, \gamma_2, \ldots \gamma_N\}$. 
%The ${\bf \Gamma}$ vector can serve as a fingerprint for both the
%calibration SNe and the candidate objects from a science sample. 
As a first-order approximation, the bias correction function for model
$M_1$ would be: 

\begin{equation}
  \beta_1(\bf \Gamma) = b_{11}\,\gamma_1 + b_{12}\,\gamma_2 + ... + b_{1N}\,\gamma_N
  \label{eqn:beta1}
\end{equation}

\noindent Here we have introduced the parameters $b_{jk}$, which are
two-dimensional constants that carry units of $z$ and \mue.\footnote{
  They could just as well include \Av\ and \tpk, but we have assumed
  that those are being marginalized out as nuisance parameters. }
If we have a candidate that is a 100\% match to model $M_1$ then
$\gamma_1=1$ and the
$b_{11}$ parameter by itself would define the necessary bias
correction.   If instead the candidate has multiple non-zero $\gamma$
values  for model $M_1$ would be a weighted sum of the $b_{jk}$
parameters. 
 
Each template $M_j$ needs its own scalable bias correction function
$\beta_j$. We can describe the complete set of bias correction
functions for all N templates in matrix form: 

\begin{equation}
\left[ \begin{array}{c} \beta_1 \\ \beta_2 \\ \vdots \\
    \beta_N \end{array} \right] = 
\left[ \begin{array}{cccc} 
    b_{11} & b_{12} & \cdots & b_{1N} \\ 
    b_{21} & b_{22} & \cdots & b_{2N} \\ 
    \vdots \\ 
    b_{N1} & b_{N2} & \cdots & b_{NN} \\ 
\end{array} \right]
\cdot
\left[ \begin{array}{c} 
    \gamma_{1} \\ \gamma_{2} \\ \vdots \\ \gamma_{N} \\ 
\end{array} \right]
\label{eqn:beta}
\end{equation}

\noindent From this setup we can determine the appropriate values for
the $b_{jk}$ parameters using a large calibration set of SNe with good
light curves and spectroscopic redshifts.  For each calibration SN, we
would build up the single-template membership functions, and integrate
them using Eq.~\ref{eqn:gammaj}.  We would then measure the distance
from the $g(z,\mue|M_j)$ peak to the SN's true location in order to
get a $\beta_j$ value for each template. Every individual calibration
SN thus provides an estimate for all components of the {\bf$\beta$}
and ${\bf\Gamma}$ vectors.  If the set of calibration SNe contains at
least as many SNe as the template library, then the system of N
equations in Equation~\ref{eqn:beta} can be solved for the optimal set
of $b_{jk}$ values.  One possible source for calibration SNe would be
to use a low-redshift Hubble flow training set
\citep[e.g.][]{Jha:1999} comprised of SNe that are far enough away to
have small peculiar velocity errors ($cz\gtrsim 2500\,km\,s^{-1}$) but
near enough that non-linear cosmological effects are negligible
($cz\lesssim 30,000\,km\,s^{-1}$).
%Incorporating training set objects from higher redshifts (such as the 
%SDSS and SNLS SNe in our verification data) would provide
%optical light curve templates that constrain the rest-frame
%ultraviolet light.  This would be a significant advantage, because the
%rest-frame UV is the region where \TNSNe\ show the greatest 
%heterogeneity.  With high redshift SN templates, however,
%spectroscopic redshifts alone are insufficient to define their
%locations.  The true luminosity distance of the calibration SNe, \mue,
%would need to be estimated using another light curve fitter, such as
%SALT or MLCS, or by assuming a cosmology and deriving \mue\ from the
%measured redshift. In either case, this step could introduce new
%systematic biases that would remove some of the autonomy of the SOFT
%method.  

This proposal is very simplistic, and may in fact
be insufficient to account for the complexities of a real bias
function.  It is quite possible that the biases from individual
templates are too inconsistent to allow for such a simple linear
decomposition.  One would need to see whether a consistent solution of
Eq.~\ref{eqn:beta} is possible, and whether the application of a bias
correction vector is actually able to improve the results for
a verification data set, such as the SDSS and SNLS SNe used here.

\subsection{Peer to Peer Cosmology}
\label{sec:PeerToPeerCosmology}

An alternative method for removing location
parameter biases is to use an approach that is based on selective
comparisons.  
The  SOFT method is founded on the assumption that two
objects with very similar physical parameters 
\bfPhi\ will necessarily have very similar light curve shapes.  
The biases that we are trying to remove arise primarily 
due to light curve shape mismatches between the template and the
candidate SN.   Thus, if we could identify a group of SNe that have
very similar \bfPhi\ values, then we would expect a direct comparison
between group members to be largely free of bias.  

One possibility would be to select candidate SNe that have very close
matches in the template library.  For example, we could pick out only
those candidate objects that have a very strong match with a single
template by elevating our template selection threshold to 
$\gamma>90\%$ or $95\%$.   
However, with only a few dozen
templates in the library most SN candidates will not have a 90\%
match with any template, so the available sample size would be
severely reduced. 

A modification of this approach could overcome the limitations of a
sparse template library by leveraging the very large SN data sets to
come from future surveys.  Suppose that a candidate SN~X is observed
at an unknown location \bftheta$_X$ and we compare it against our N
templates to get a vector of integrated membership grades ${\bf
  \Gamma_X}=(\gamma_{X1},\gamma_{X2},\ldots \gamma_{XN})$.  A second
candidate SN~Y exists at a different location \bftheta$_Y$ and has a
different set of membership grades ${\bf \Gamma}_Y$.  If these two
objects have very similar physical characteristics, 
$\bfPhi_X\approx\bfPhi_Y$, then we should expect that the same
templates that are good matches for SN~X will also be similar to SN~Y,
so that ${\bf \Gamma}_X \approx {\bf \Gamma}_Y$.

This provides a straightforward means for collecting a group of
similar SNe, which we will call a ``peer group.''  We compare all the
available candidates against the template models, and then group
together candidates that have a very similar pattern of membership
grades ${\bf \Gamma}$.  The members of this peer group generally will
not have a perfect match in the template library, so no single
template can provide an unbiased estimate for any of the group
members.  However, we do expect that any member of this peer group
could provide an unbiased estimate for the location parameters of any
other member in the group.  All that is needed is to redefine the
candidate SN members to take the place of the template library.

Recognizing that absolute distances are not necessary
for cosmology, let us reduce the location vectors for our two peer SNe
X and Y down to a pair of three-dimensional sub-vectors
$\bfvartheta_X=(z_X,A_{V,X},t_{pk,X})$ and
$\bfvartheta_Y=(z_Y,A_{V,Y},t_{pk,Y})$.
The procedure for executing a peer to peer comparison is as follows:
\begin{enumerate}
\item Fit a smooth spline curve to the photometric data of 
  SN~X, allowing us to interpolate to any point in time.  
\item Make a guess for $\bfvartheta_X=(z_X,A_{V,X},t_{pk,X})$  
\item For every spectrum in the SOFT spectral library (Paper I,
  \S2.2), apply the $z_X$ redshift, add reddening according to
  $A_{V,X}$, and shift in time to match $t_{pk,X}$.
\item Iteratively warp the adjusted spectra to force agreement with
  the observed $griz$ photometry of SN~X.
\item Remove the effects of the assumed $\bfvartheta_X$ from the
  warped spectra 
  (i.e. apply a reverse time shift, dereddening, and a blue-shift) 
  to get a rest-frame spectrophotometric model for SN~X.
\item Repeat steps 1-5 for SN~Y to get a rest-frame spectrophotometric
  model of SN~Y under the assumption of a location 
  $\bfvartheta_Y=(z_Y,A_{V,Y},t_{pk,Y})$.
\end{enumerate}

At the end of this process we will have a pair of rest-frame
spectrophotometric models $f_X(\lambda,t)$ and $f_Y(\lambda,t)$ giving
the flux as a function of wavelength and time for both SN~X
and SN~Y, at the assumed locations $\bfvartheta_X$ and 
$\bfvartheta_Y$.   We now need to consider the relative luminosity
distances for these two objects. 
Assuming that membership in the same peer group indicates that
these two objects are physically almost identical, then they should
have nearly the same intrinsic luminosity.  This means that their
relative luminosity distances $D_{L,X}$ and $D_{L,Y}$ are encoded in
the flux ratio:  

\begin{equation}
\mathcal{R}_f = \frac{f_{X}}{f_Y} = \left(\frac{D_{L,Y}}{D_{L,X}}\right)^2
\label{eqn:Rf}
\end{equation}

With the addition of $\mathcal{R}_f$ we now have seven free parameters
that define our rest-frame models for both objects. If the two SNe are
in fact physically similar, and if we have guessed correctly about
$\bfvartheta_X$, $\bfvartheta_Y$ and  $\mathcal{R}_f$, then we
should expect to see the same flux values as a function of wavelength
and time: $f_X(\lambda,t) = \mathcal{R}_f \cdot f_Y(\lambda, t)$.  This
means that we can compute a posterior likelihood for each set of model
parameters in the same manner that was used for
Eq.~\ref{eqn:p(D|theta,Mj)}:

\begin{equation}
  \begin{array}{rl}
    p(\bfD|&\bfvartheta_X,\bfvartheta_Y,\mathcal{R}_f)  
    =  \prod\limits_{i=1}^N  \frac{1}{\sqrt{2\pi}\sigma_i} \times \\
    & \mbox{exp}\left(\frac{-(f_X(t_i,\lambda_i)-\mathcal{R}_f\cdot f_{Y}(t_i,\lambda_i))^2}{2\sigma_i^2}\right) 
    \end{array}
\label{eqn:p(D|varthetaX,varthetaY,Rf)}
\end{equation}

\noindent Here we are computing a flux difference $f_X-R_ff_Y$ at N
points in time and wavelength.\footnote{ 
  In theory, N could be an exceedingly
  large number, as the rest-frame models for SN~X and SN~Y could
  encompass several dozen points in time and several hundred
  distinct wavelengths. To reduce the computational complexity, it
  would be advisable to integrate the spectrophotometric models into a
  handful of broad bandpasses in the visual wavelength range (these
  could be arbitrary box-car functions or real filter transmission
  curves), and to select a few representative time points for
  comparison (perhaps t=-5,0,+10,+30 days from the peak).}
The flux uncertainty term, $\sigma_i$ could be computed as a quadratic
sum of uncertainties from the SN~X and SN~Y models, accounting for
the true observational errors for both SNe as well as additional error
introduced by the spline fitting and spectral warping steps described
above. 

The posterior probability of Eq.~\ref{eqn:p(D|varthetaX,varthetaY,Rf)}
can now be treated in the same manner as the probabilities derived for
the template-based SOFT comparisons in
Section \ref{sec:ParameterEstimation}.  After setting 
priors for all the components of $\bfvartheta_X$, $\bfvartheta_Y$ and
$\mathcal{R}_f$, we can apply Bayes' theorem as in
Eq.~\ref{eqn:p(theta|D,Mj)}.  The resulting probability distribution
can be marginalized over nuisance parameters to get a
three-dimensional function in ($z_X$,$z_Y$,$\mathcal{R}_f$) space,
which provides a direct constraint on any cosmological model. 

There are several advantages to be gained by adopting this type of
strategy for measuring cosmological parameters without spectroscopy.  
In addition to the bias minimization benefit, using peer to peer
comparisons could provide a more robust model for the rest-frame
ultraviolet light. For example, the rest-frame U band light from a SN
at redshift z$>$0.2 will be observed in visual bandpasses (V or g). 
Thus, the g-band light curve of a z=0.2 template drawn from the
Pan-STARRS survey is sufficient for comparison to the r-band light
curve of a z=0.5 SN from the same survey.  This approach also 
provides a natural set of internal consistency checks.  As SN surveys
begin to collect thousands of \TNSN\ light curves, it may be possible
to define peer groups with 10 or 20 members that all have very
strongly similar light curve shapes.   Each direct comparison
between two objects in a peer group will provide an estimate of the
redshifts for both objects.  The consistency of redshift estimates for
each object in the group provides a useful verification that the peer
group is indeed defining a homogeneous set.

\subsection{Spectral Library Bias}
\label{sec:SpectralLibraryBias}

In addition to the biases of a sparse template library discussed
above, SOFT may also suffer from systematic biases introduced by the
spectral  library.  We have used the spectral templates of
\citet{Hsiao:2007} for normal Type Ia SNe, and those of
\citet{Nugent:2002} to describe overluminous 91T-like SNe and
underluminous 91bg-like SNe. 
It has been suggested that the
spectral properties of TN SNe may evolve with redshift
\citep{Blondin:2006,Bronder:2008,Foley:2008,Sullivan:2009}, 
in which case the use of these spectral templates based primarily on
low-z SNe becomes inappropriate.   This problem is particularly acute
in the ultraviolet, where there can be substantial spectral variation
\citep{Ellis:2008}, and where it is especially difficult to collect
a local sample \citep{Foley:2008a}.  The SALT2 light curve fitter
addresses the UV problem by incorporating high redshift spectra to
define the rest-frame UV \citep{Guy:2007}.  Extending the SOFT
spectral library to avoid possible bias will be especially important
when SOFT is applied for parameter estimation of TN SNe with
z$>0.8$, where the rest-frame UV dominates the visible bandpasses.

\section{Summary}
\label{sec:Summary}

In Paper I we introduced the SOFT program, using a set of fixed-shape
light curve templates and the framework of fuzzy set theory for
combining results from multiple templates.  We applied SOFT as a SN
classification tool in Paper I, and in this companion paper we have
shown how SOFT can also be used to estimate parameters of cosmological
interest: redshift $z$ and luminosity distance \mue. The SOFT method
is distinct from light curve fitters such as MLCS and SALT in that it
does not describe the light curve shape with a parameterized model.
Given this fundamental distinction, SOFT may provide a valuable
addition to the parametric fitters in that it could have smaller $-$
or at least {\em different} $-$ systematic uncertainties.

Using Type Ia SN light curves from the SDSS and SNLS data sets, we
have performed a set of verification tests to demonstrate the accuracy
of SOFT.  We found that the SOFT redshift estimates are comparable to
optical photo-z measurements for SDSS galaxies: both methods yield
$\delta z=z_{SOFT}-z_{spec}$ residuals with an RMS scatter of
approximately 0.05. Applying SOFT to derive (z,\mue) coordinate pairs
for a joint SDSS$+$SNLS sample, we find an RMS scatter around the
$\Lambda$CDM model of 0.18 mags in \mue\ when including spectroscopic
redshift priors. When SOFT analyzes the SDSS$+$SNLS sample with no
spectroscopic information at all, we find the RMS scatter of Hubble
residuals increases to 0.37 mags.

To investigate the near future of SN cosmology, we considered the
degree to which a variable DE model can be constrained by larger SN
samples containing thousands of light curves but very little
spectroscopic follow-up.  Using  a bootstrap Monte Carlo
approach, we simulated the distance and redshift estimates that might be
obtained by SOFT for 9 different survey structures.  We find that SOFT
should be able to improve the DETF FoM by a factor of 2--3 with each new
generation of SN surveys.  

The SOFT program has not yet realized the full potential of its fuzzy
logic approach, but we have outlined several pathways for improving
the method and limiting systematic biases. We have proposed a
straightforward method for using a training set to calibrate the
template library, and have also presented a ``Peer to Peer Cosmology''
approach in which SOFT can identify groups of similar SNe and do a
direct comparison to determine their redshifts and relative distances.
This latter method may be especially applicable in upcoming SN surveys
such as Pan-STARRS and LSST, which will have an abundance of
well-sampled multi-color light curves, but comparatively little
spectroscopic followup.

{\bf Acknowledgments:}
We would like to thank the anonymous referee for a thorough and
critical reading of this work, which led to substantial improvements.

\bibliographystyle{apj}
\bibliography{bibdesk}

\end{document}